# Random walks in disordered lattice, CTRW, memory and dipole transport

F.S. Dzheparov

National Research Center "Kurchatov Institute", ITEP, Moscow 117258, Russia,

*Application of CTRW to dipole hopping transport is considered. Correct versions of derivation of the CTRW-equations are presented. Existence of different forms of memory kernels is demonstrated. Correction of Scher-Lax memory kernel within geometrical memory approach is fulfilled in accordance with leading terms of concentration expansion. Approximate solution for autocorrelation function is constructed. Modern state of numerical simulation and experimental measurements of autocorrelation function in nuclear polarization delocalization is described.*

## 1. Introduction

Random walks in disordered lattice are described by the equations

$$\frac{\partial}{\partial t} p_{i0} = -\sum_j (w_{ji} p_{i0} - w_{ij} p_{j0}), \quad p_{i0}(t=0) = \delta_{i0}, \quad (1)$$

where $p_{i0}(t)$ is the "probability" to find an "excitation" at position $\mathbf{r}_i$, if it started at $\mathbf{r}_0 = \mathbf{0}$ at $t = 0$ and $\delta_{ij}$ is the Kronecker's symbol. Here $w_{ji}$ is the transition rate for transfer from $\mathbf{r}_i$ to $\mathbf{r}_j$. The problem of disordered sites will be considered below, when positions $\mathbf{r}_j$ are statically and randomly distributed on sites of regular lattice, $w_{ji}$ depends on $\mathbf{r}_{ij} = \mathbf{r}_i - \mathbf{r}_j$, and observables are directly related with solution $p_{i0}(t)$, averaged on all possible positions $\{\mathbf{r}_j\}$. Occupation number representation admits to rewrite Eqs.(1) as [1]

$$\frac{\partial}{\partial t} \tilde{P}_{\mathbf{x}0} = -\sum_{\mathbf{z}} n_{\mathbf{x}} n_{\mathbf{z}} \left( w_{\mathbf{z}\mathbf{x}} \tilde{P}_{\mathbf{x}0} - w_{\mathbf{x}\mathbf{z}} \tilde{P}_{\mathbf{z}0} \right) = -\left( A^{(0)} \tilde{P} \right)_{\mathbf{x}0}, \quad (2)$$

or

$$\frac{\partial}{\partial t} \tilde{P}_{\mathbf{x}0} = -\sum_{\mathbf{z}} \left( n_{\mathbf{z}} w_{\mathbf{z}\mathbf{x}} \tilde{P}_{\mathbf{x}0} - n_{\mathbf{x}} w_{\mathbf{x}\mathbf{z}} \tilde{P}_{\mathbf{z}0} \right) = -\left( A^{(1)} \tilde{P} \right)_{\mathbf{x}0}, \quad (3)$$

where the propagator $\tilde{P}_{\mathbf{x}0}$ gives the probability to find the excitation at lattice site $\mathbf{x}$, when initially it was at the site $\mathbf{0}$, and $w_{\mathbf{z}\mathbf{x}} = w_{ij}(\mathbf{r}_i = \mathbf{z}, \mathbf{r}_j = \mathbf{x})$. Here $n_{\mathbf{r}}$ is occupation number of the site $\mathbf{r}$ by a donor ($n_{\mathbf{r}} = 1(0)$ if the site $\mathbf{r}$ is (not) occupied by the donor), while the donor is an impurity, which can carry the excitation. Eqs. (3) are equivalent to Eqs. (2) because $n_{\mathbf{x}} \tilde{P}_{\mathbf{x}0} = \tilde{P}_{\mathbf{x}0}$, while equivalence of Eqs. (1) and (2) is evident, if we omit in (2) all empty sites for which $n_{\mathbf{x}} = 0$ and, consequently, $\tilde{P}_{\mathbf{x}0} = 0$.

Solutions of these equations are of the form $\tilde{P}_{\mathbf{x}0} = \left( \exp(-At) \right)_{\mathbf{x}0} n_0 / c$, where the operator $A = A^{(0)}$ or $A = A^{(1)}$, and the initial condition

$$\tilde{P}_{\mathbf{x}0}(t=0) = n_{\mathbf{x}} \delta_{\mathbf{x}0} / c \quad (4)$$

ensures that the excitation can be placed at site $\mathbf{0}$ if it is occupied by a donor and produces evident normalization

$$\sum_{\mathbf{x}} P_{\mathbf{x0}}(t) = 1.$$

Occupations of different sites are assumed as independent and having no dependence on the $\mathbf{r}$ (with small probability of occupation $c \ll 1$ as a rule):

$$\langle n_{\mathbf{r}} \rangle = c, \quad \langle n_{\mathbf{r}} n_{\mathbf{x}} \rangle = c\delta_{\mathbf{rx}} + c^2(1 - \delta_{\mathbf{rx}}), \quad \langle \prod_{j=1}^{'m} n_{\mathbf{r}_j} \rangle = \prod_{j=1}^{'m} \langle n_{\mathbf{r}_j} \rangle = c^m.$$

Here $\langle \cdots \rangle$ is averaging on all possible donors positions in infinite sample. All $\mathbf{r}_j$ are different in the last relation. Coincidence in indexes can be treated using the identity $n_{\mathbf{r}}^2 = n_{\mathbf{r}}$.

The problem consists in calculation of the observable propagator

$$P_{\mathbf{x0}}(t) = \langle \tilde{P}_{\mathbf{x0}}(t) \rangle. \tag{5}$$

Most important experimental realizations of the process (1) are hopping conductivity, for which the CTRW theory of Scher and Lax [2] was developed, Foerster electronic energy transfer [3] and spin-polarization transfer [1]. The simplest transition rates are of the form

$$w_{\mathbf{z} \neq \mathbf{x}} = w_0 \exp(-|\mathbf{z} - \mathbf{x}|/r_0), \quad w_{\mathbf{xx}} = 0 \tag{6}$$

for conductivity problem, and

$$w_{\mathbf{z} \neq \mathbf{x}} = w_0 r_0^6 / (\mathbf{z} - \mathbf{x})^6, \quad w_{\mathbf{xx}} = 0 \tag{7}$$

for dipole transitions in electronic energy and spin-polarization transfers. Here $w_0$ defines the transition rate at distance $r_0$ between donors. More advanced representations for transition rates can be found in special literature, see, for example, [4] for spin polarization transfer. It should be noted that in the problem of spin-polarization transfers the propagator $P_{\mathbf{x0}}(t)$ represents polarization (instead of probability) of a spin at the site $\mathbf{x}$, when initially polarization was localized at the site $\mathbf{0}$.

The problems of calculation of the propagator (5) or memory kernels $N_{\mathbf{xz}}(t)$ in corresponding master equation

$$\frac{\partial}{\partial t} P_{\mathbf{x0}}(t) = -\int_0^t dt' \sum_{\mathbf{z}} [N_{\mathbf{zx}}(t') P_{\mathbf{x0}}(t-t') - N_{\mathbf{xz}}(t') P_{\mathbf{z0}}(t-t')], \quad P_{\mathbf{x0}}(t=0) = \delta_{\mathbf{x0}} \tag{8}$$

belong to the most complex problems of modern statistical mechanics and they have not adequate analytical solution up to now. Nonseparable version of CTRW theory, developed by Scher and Lax in two articles [2], produced an integral representation for approximate solution of the problem and extended previously developed separable CTRW of Montroll and Weiss [5] (which operated with kernels of the form: $N_{\mathbf{zx}}(t) = X_{\mathbf{zx}} \cdot Y(t)$).

The Scher-Lax theory was directed on calculation of frequency dependent diffusion (or conductivity), which are defined by

$$\langle \mathbf{x}^2(t) \rangle = \sum_{\mathbf{x}} \mathbf{x}^2 P_{\mathbf{x0}}(t),$$

and it produced important progress in the field. But the theory was ineffective in description of other important value, survival probability (or autocorrelation function) $P_{00}(t)$, which is directly measurable in fine optical [6,7] and beta-NMR [4,8] studies. Other important property - conceptual foundation of the relations between "microscopic" equations (1)-(3) and master equations (8) still unrecognized by absolute majority of workers, that is clearly seen from legend about the derivation of the master equation which exists 37 years and passed in erroneous form throw many reviews, see for example [9, ch.5] and [10].

The aims of this article consists in explaining of nonseparable version of CTRW, constructed by me and my colleagues for qualitatively correct description of the autocorrelator $P_{00}(t)$ for problems of dipole hopping transport, in improvement of some details of the theory, in short description of corresponding modern numerical and experimental results and in concentrated clarifying of those connections of the master equations (8) with primary equations (1)-(3), which are fundamentally important for correct calculation of the $P_{00}(t)$ and for the problem of random walks on disordered sites as a hole.

**2. Reformulation and generalization of the CTRW theory**
The method of approximate constructions of the propagator $P_{x0}(t)$, developed in Refs. [2], was reformulated in [11,12] (see also [13,14] for different approach) basing on the following assumptions:
a) the first term in square brackets in (8) is the rate of polarization outflow from $\mathbf{x}$ to $\mathbf{z}$, while the second term is the rate of inflow from $\mathbf{z}$ to $\mathbf{x}$;
b) inflow rate from $\mathbf{z}$ to $\mathbf{x}$ does not depend on how the polarization reached $\mathbf{z}$.
The processes of inflow to the site and outflow from it are now separated, and to determine the kernels $N_{zx}(t)$ we can treat the simple case in which an exact averaging can be carried out. For this purpose, we choose the process of the outflow of polarization from an arbitrary site in exactly solvable problem

$$\frac{\partial}{\partial t}\tilde{F}_{xx} = -\sum_z n_z w_{zx}\tilde{F}_{xx}, \quad \frac{\partial}{\partial t}\tilde{F}_{z\neq x} = n_z w_{zx}\tilde{F}_{xx}, \qquad (9)$$

$$\tilde{F}_{zx}(t=t_0) = n_z \delta_{zx}/c, \quad \tilde{F}_{zx}(t<t_0) = 0.$$

According to the assumptions above, the average $F_{zx}(t) = \langle \tilde{F}_{zx}(t)\rangle$ must satisfy the equations

$$\frac{\partial}{\partial t}F_{xx}(t) = -\int_0^t dt'\sum_z N_{zx}(t')F_{x0}(t-t'), \quad \frac{\partial}{\partial t}F_{z\neq x}(t) = \int_0^t dt' N_{zx}(t')F_{xx}(t-t'), \qquad (10)$$

$$F_{zx}(t=t_0) = \delta_{zx}, \quad F_{zx}(t<t_0) = 0.$$

As a result, the equation for the kernel $N_{zx}(t)$ receives the form

$$\int_0^t dt' N_{zx}(t')\left\langle \exp\left(-\sum_q n_q w_{qx}(t-t')\right)\right\rangle = \left\langle n_z w_{zx}\exp\left(-\sum_q n_q w_{qx}t\right)\right\rangle \qquad (11)$$

with

$$\left\langle \exp\left(-\sum_q n_q w_{qx}t\right)\right\rangle = \exp\left(\sum_q \ln\left(1-c\left(1-e^{-w_{qx}t}\right)\right)\right), \qquad (12)$$

$$\left\langle n_z w_{zx}\exp\left(-\sum_q n_q w_{qx}t\right)\right\rangle = cw_{zx}e^{-w_{zx}t}\exp\left(\sum_{q\neq z}\ln\left(1-c\left(1-e^{-w_{qx}t}\right)\right)\right). \qquad (13)$$

The Eqs. (8) and (11) give qualitatively satisfactory description of those properties of the hopping conductivity and delocalization of excitations, which are related with $\langle \mathbf{x}^2(t)\rangle = \sum_{\mathbf{x}}\mathbf{x}^2 P_{x0}(t)$, but, as it is demonstrated in next section, they are erroneous in the description of the autocorrelator $P_{00}(t)$, which is directly measurable in fine optical [3, 6, 7] and beta-NMR [4, 8] experiments.

**3. Main weakness of the Scher-Lax CTRW theory.**
In order to sharpen the problem we can consider the continuum media approximation, when impurity concentration $c \to 0$ and lattice prime cell volume $\Omega \to 0$ at a fixed value of impurity density $n = c/\Omega$. In the continuum media approximation the Eqs. (8) receive the form

$$\frac{\partial}{\partial t}P(\mathbf{x},t\,|\,\mathbf{0}) = -\int_0^t dt'\int d^3z\left[N_{zx}(t')P(\mathbf{x},t-t'\,|\,\mathbf{0}) - N_{xz}(t')P(\mathbf{z},t-t'\,|\,\mathbf{0})\right] =$$

$$= -\int_0^t dt' \int d^3z\, Z_{xz}(t') P(\mathbf{z}, t-t' | \mathbf{0}), \quad P(\mathbf{z}, t=0 | \mathbf{0}) = \delta(\mathbf{x}), \tag{14}$$

where $\delta(\mathbf{x})$ is Dirac's delta-function, the probability density $P(\mathbf{x}, t | \mathbf{0}) = P_{\mathbf{x}0}(t)/\Omega$ has normalization $\int d^3 x P(\mathbf{x}, t | \mathbf{0}) = 1$ and the kernels can be written as

$$N_{zx}(\lambda) = \int_0^\infty dt\, e^{-\lambda t} N_{zx}(t) = n w_{zx} Q(\lambda + w_{zx}) / Q(\lambda), \tag{15}$$

$$Q(\lambda) = \int_0^\infty dt\, e^{-\lambda t} Q(t) = \left(\lambda + \int d^3 z N_{zx}(\lambda)\right)^{-1},$$

$$Q(t) = \left\langle \exp\left(-\sum_\mathbf{q} n_\mathbf{q} w_{\mathbf{qx}} t\right)\right\rangle = \exp\left(-n \int d^3 q \left(1 - e^{-w_{\mathbf{qx}} t}\right)\right) = e^{-(\beta t)^{1/2}}. \tag{16}$$

The Foerster constant $\beta = (16/9)\pi^3 n^2 w_0 r_0^6$ is defined by the last equality in (16). Here and below we apply the same symbols for time dependent functions and for their Laplace transformations distinguishing them by the argument.

The master equation (14) can be rewritten in the Laplace representation as

$$P(\mathbf{x}, \lambda | \mathbf{0}) = Q(\lambda)\left(\delta(\mathbf{x}) + \int d^3 z N_{xz}(\lambda) P(\mathbf{z}, \lambda | \mathbf{0})\right) = P^{(s)}(\mathbf{x}, \lambda | \mathbf{0}) + P^{(r)}(\mathbf{x}, \lambda | \mathbf{0}), \tag{17}$$

or, in the time representation

$$P(\mathbf{x}, t | \mathbf{0}) = P^{(s)}(\mathbf{x}, t | \mathbf{0}) + P^{(r)}(\mathbf{x}, t | \mathbf{0}), \quad P^{(s)}(\mathbf{x}, t | \mathbf{0}) = Q(t)\delta(\mathbf{x}). \tag{18}$$

The solution (15) indicates that $N_{xz}(\lambda)$ is a smooth function of $|\mathbf{x} - \mathbf{z}|$, therefore Eqs. (17) and (18) separate the propagator on singular $P^{(s)}(\mathbf{x}, t | \mathbf{0})$ and regular $P^{(r)}(\mathbf{x}, t | \mathbf{0})$ parts near $x = 0$. At that, the singular part is defined only by the singular part of the memory kernel $Z_{xz}^{(s)}(t) = \delta(\mathbf{x} - \mathbf{z}) \int d^3 q N_{\mathbf{qx}}(t)$ of the Eq. (14).

Simple analysis indicates, that for long time

$$\langle \mathbf{x}^2(t) \rangle = \int d^3 x\, x^2 P(\mathbf{x}, t | \mathbf{0}) \sim n^{-2/3} \beta t, \tag{19}$$

that is in agreement with scaling arguments [1,15] and expected for diffusion long time asymptotics. But the solution $P^{(s)}(\mathbf{x}, t | \mathbf{0}) = F_0(t)\delta(\mathbf{x})$ with $F_0(t) = Q(t)$ is incorrect [1, 7, 11, 12]. It decays with time exponentially, while more slow behavior, diffusion like as $F_0(t) \sim (\beta t)^{-3/2}$ should be expected, because $F_0(t)$ is the survival probability, and it should be of the order of probability $F_1(t)$ to find the excitation on a donor, placed near the origin, that is

$$F_1(t \to \infty) \sim \int d^3 x\, \vartheta(x < n^{-1/3}) P_{\mathbf{x}0}^{(r)}(t \to \infty) \sim \frac{1}{n} \cdot \frac{1}{\langle x^2(t) \rangle^{3/2}} \sim (\beta t)^{-3/2}. \tag{20}$$

Here Heaviside's function $\vartheta(x)$ is applied.

The relations (17) and (18) indicate, that correct $F_0(t \to \infty)$ can be obtained if the singular part of the memory kernel $Z_{xz}^{(s)}(t) = \delta(\mathbf{x} - \mathbf{z}) Z_D(t)$ has correct long time tail $Z_D(t \to \infty) = -F_0(t) / \int_0^\infty dt' F_0(t') \sim -(\beta t)^{-3/2}$ [12]. This conclusion contradicts to main strategy of application of the memory functions method when reasonable approximation for short-term memory kernels produces satisfactory long time behavior for the solution of master equation (that was fulfilled in the relation (19) for example). Therefore, we should look for justification of the applicability of the memory function method (i.e. Eq. (8)) and for modification of the memory kernels.

**4. Correction of the CTRW theory**
The justification of the applicability of the memory function method can be based on derivation of the Eq. (8) applying the Nakajima-Zwanzig projection operator technique to Eqs. (2) or (3). Similar attempt was undertaken in Ref. [16] for Eqs. (2) choosing the projection operator $\hat{\pi}$ as simple averaging $\hat{\pi}A = \langle A \rangle$ for any $A$. But initial condition was applied in incorrect form $\tilde{P}_{x0}(t=0) = \delta_{x0}$, therefore all equations, derived in [16], are incorrect [17]. Nevertheless, authors of reviews [9] and [10], as well as more than three hundreds other workers insist that the Ref. [16] gave convincing derivation of the master equation (8) ignoring the criticism of the Ref. [17]. It should be noted nevertheless that similar derivation can be fulfilled both for Eqs. (2) and (3) with correct initial condition $\tilde{P}_{x0}(t=0) = n_x \delta_{x0} / c$ applying other projection operator

$$\hat{\pi}A_x = \frac{n_x}{c}\langle A_x \rangle. \qquad (21)$$

Unfortunately, the results of these derivations produce no indication on the way to improve the singular part of memory kernels.

More constructive approach was realized in Refs. [11], [12], see also [18]. Correct initial condition $\tilde{P}_{x0}(t=0) = n_x \delta_{x0} / c$ separates only those solutions of the Eqs. (2) and (3), for which the lattice site $\mathbf{r} = \mathbf{0}$ is occupied by a donor, because the identity $n_x R(n_x) = n_x R(1)$ is valid for any reasonable function $R(n_x)$. Therefore,

$$P_{x0}(t) = \left\langle \left( \exp(-At) \right)_{x0} \frac{n_0}{c} \right\rangle = \left\langle \left( \exp(-At) \right)_{x0} \right\rangle_0, \qquad (22)$$

where $\langle \cdots \rangle_0$ means averaging on occupation numbers with the condition $n_0 = 1$. Therefore, we can apply the projector $\hat{\pi} = \pi_0$ acting as

$$\pi_0 B = \langle B \rangle_0 \qquad (23)$$

for any $B$. Standard transformations produce the Eqs. (8) and (14) again, but the memory kernel $N_{zx}(t) = N^{(0)}_{zx}(t)$ depends both on $\mathbf{z} - \mathbf{x}$ and $\mathbf{x} = \mathbf{x} - \mathbf{0}$, while the propagator $P_{x0}(t)$ depends on $\mathbf{x}$ only. This new type of memory (geometrical, contrary do dynamical one, which depends on $\mathbf{z} - \mathbf{x}$ only) is much less comfortable for calculations, because the matrix $N^{(0)}_{zx}(t)$ can not be diagonalized by the Fourier transformation. Nevertheless [11,12], equations for the memory kernels can be constructed following the derivation (9)-(13) with the result:

$$\int_0^t dt' N^{(0)}_{zx}(t') Q^0_x(t-t') = \frac{cw_{zx}\exp(-w_{zx}t)}{1+c(\exp(-w_{zx}t)-1)} Q^0_x(t), \quad \mathbf{z} \neq \mathbf{0}, \qquad (24)$$

$$N^{(0)}_{0x}(t) = \frac{w_{0x}}{cw_{x0}} N^{(0)}_{x0}(t), \quad Q^0_x(t) = \left\langle \exp\left(-\sum_{z \neq 0} w_{zx} t\right) \right\rangle = \prod_{z \neq 0}\left(1 + c\left(e^{-w_{zx}t} - 1\right)\right).$$

Details of the derivation (with taking into account additional exactly solvable model) can be found in Ref. [12].

As a result, the Eqs. (8) with short-term memory (24) produce qualitatively well-formed solution $P_{x0}(t)$ for all $\mathbf{x}$ and $t$, and the solution is correct up to terms $\sim c^1$ [11,12]. It should be noted, that last property was not fulfilled in [2].

**5. Approximate solution to corrected CTRW equations**

Analytical solution of the Eqs. (8) with kernels (24) is absent. Therefore approximate solution for $P_{00}(t)$ was constructed by matching the short- and long-time asymptotics. Analysis of long-time asymptotics [11,12] indicates, that for dipole transport

$$P_{00}(\beta t \to \infty) = \frac{1}{c} G_{00}(t)\left(1 + O\left((\beta t)^{-1}\right)\right), \tag{25}$$

where $G_{x0}(t)$ is the solution of Eqs. (8) with kernels (11). It can be received using the lattice Fourier transformation. For $\beta t \to \infty$

$$G_{00}(t) = \frac{\Omega}{(2\pi)^3} \int_B d^3k \exp\left(-(N(\mathbf{0}) - N(\mathbf{k}))t\right)\left(1 + O\left((\beta t)^{-1}\right)\right), \tag{26}$$

where integration is limited by the Brillouin zone and $N(\mathbf{k}) = \sum_{\mathbf{x}} e^{-i\mathbf{k}\mathbf{x}} \int_0^\infty dt N_{x0}(t)$. The asymptotics $G_{00}(t \to \infty)$, according to the Laplace's method, is defined by $N(\mathbf{k} \to 0)$. In simple lattice with cubic symmetry and for dipole transition rates (7) we have

$$N(\mathbf{0}) - N(\mathbf{k}) = Dk^2 - \sigma k^3 + O(k^4). \tag{27}$$

Simple derivation of this expansion can be found in Ref. [19]. The diffusion coefficient $D$ is model dependent, and its value for Scher-Lax theory in continuum media approximation is

$$D_{SL} = n^{4/3} w_0 r_0^6 \frac{\pi^{3/2} 2^{2/3}}{3^{7/3}} \Gamma(1/6)\Gamma(5/3) = \frac{\kappa_{SL}}{6} \beta n^{-2/3}, \tag{28}$$

where $\kappa_{SL} = 0.3725$, while $\sigma = \frac{\pi^2 n}{12} w_0 r_0^6$ is model independent and it is defined by the dipole long ranging exclusively. For example, if we will assume in Eqs. (15) and (16) that $Q_a(t) = \exp\left(-(\gamma t)^\alpha\right)$ with arbitrary $\alpha > 0$ and $\gamma > 0$ instead of $Q(t) = \exp\left(-(\beta t)^{1/2}\right)$, prescribed by the relation (16), then we will receive other value for the diffusion coefficient but the same $\sigma = \frac{\pi^2 n}{12} w_0 r_0^6$. As a result, in leading terms,

$$P_{00}(\beta t \to \infty) = \frac{1}{n(4\pi Dt)^{3/2}}\left(1 + \frac{4\sigma}{D^{3/2}(\pi t)^{1/2}}\right) = \frac{1}{(\mu\beta t)^{3/2}}\left(1 + \frac{\varphi}{(\mu\beta t)^{1/2}}\right), \tag{29}$$

where $\mu = 0.7801$ and $\varphi = 1.923$.

Short time asymptotics was received in Refs. [1], [3], [17] and [20]:

$$P_{00}(\beta t < 1) = 1 - (\beta t / 2)^{1/2} + d\beta t + \cdots, \tag{30}$$

where the coefficient $d$ was calculated in [17] basing on exact expansion of the propagator in powers $c^m$ of the concentration.

These results allow to construct the approximation

$$P_{00}(t) = Q(t) + \frac{1-Q(t)}{\left(\mu\beta(t+\tau)\right)^{3/2}}\left(1 + \frac{\varphi}{\left(\mu\beta(t+\tau)\right)^{1/2}}\right), \quad (31)$$

which reproduces the relation (30) up to $(\beta t)^{1/2}$ and both terms of the asymptotics (29) with $\mu\beta\tau = 3.613$ [7, 12].

The coefficients $\sigma$ and $\varphi$ prescribe specific evolution of $P_{00}(t)$ relative to its long time asymptote $P_a(t) = (\mu\beta t)^{-3/2}$. At the beginning $P_{00}(t) < P_a(t)$, but with increasing of time we have opposite relation and at $\beta t \to \infty$ they coincide. This property (reoscillation) was applied in the Ref. [7] to clarify, that, in agreement with Eq. (31), the onset of the diffusive asymptotic behavior in the kinetics of the electodipole delocalization of excitations in a disordered system of donors takes place at $P_{00}(t) < 0.03$.

Consequent numerical [21], [22] and experimental (beta-NMR) [23], [24] studies revealed, that
a) for small concentrations (i.e. for strongest disorder) the diffusion coefficient is

$D_0 = \frac{\kappa_0}{6}\beta n^{-2/3}$ with $\kappa_0 = 0.296$, this is not very far from $\kappa_{SL} = 0.3725$ in the relation (28), and

b) for $c \leq 0.1$ and for all $t$ autocorrelator $P_{00}(t)$ is of the form

$$P_{00}(t) = F_{00}(t)(1 + \chi(t)), \quad (32)$$

where $F_{00}(t)$ is defined by the relation (31) with more correct diffusion tensor (or coefficient, for isotropic transfer) and variation $\chi(t)$ is relatively small ($-0.1 < \chi(t) < 1$). The variation increases the reoscillation.

It should be noted that, according to (29), for recalculation to new value of the diffusion coefficient, the relation (31) can be written as

$$F_{00}(t) = Q(t) + \frac{1-Q(t)}{\left(\mu\beta(t+\tau)\right)^{3/2}}\left(1 + \frac{D_{SL}}{D}\cdot\frac{\varphi}{\left(\mu\beta(t+\tau)\right)^{1/2}}\right) \quad (33)$$

in order to apply the same value $\varphi$. The multiplayer $D_{SL}/D$ was forgotten in preceding studies, but it did not produced errors in description of results of numerical studies, because 1) for approximations of the numerical results the relation (32) was applied as a hole with the fitting function $\chi(t)$, and corresponding errors was compensated by $\chi(t)$, and 2) for studied values of $D$ with corresponding $\mu\beta\tau$ the relative inaccuracy in calculation of $F_{00}(t)$ never exceeded 0.05. Nevertheless we should expect that the correction will become important with increasing of accuracy of theoretical and experimental studies, because it admits to exclude the term $\sim (\beta t)^{-1/2}$ from the fitting function $\chi(t)$.

## 6. Conclusions

As a hole we see, that the version of CTRW, invented in Refs. [2] and improved in Refs. [11] and [12], produced important part of the basis for consequent quantitative understanding and experimental investigations of the problem of random walks in disordered media with dipole transitions. It should be noted, that this approach admitted for the first time to receive analytically correct diffusion long-time asymptotics for autocorrelator $P_{00}(t) \sim (\beta t)^{-3/2}$, while other methods (see, for example, Ref. [3], [20] and [25]) produced reasonable behavior at $\beta t \sim 1$, but exponential long time tail. Diffusion long time tail can be received for coarse-grained propagator, as in the Ref. [26], this is evident from the relation (20), but it produces no direct information for comparison with precise optical and beta-NMR studies.

We can state, that the CTRW theory was more successful in the description of dipole processes, than for hopping conductivity, where the percolation theory is more applicable. Indeed, CTRW and the relation (28) produces correct dependence of the diffusion coefficient $D(c) \sim c^{4/3} \sim n^{4/3}$ for dipole transport, but in the problem of hopping conductivity $D \sim \exp\left(-\kappa_p / \left(r_0 n^{1/3}\right)\right)$, where the constant $\kappa_p$ is produced by the percolation theory, while CTRW gives other parametric dependence [2,13,27].